# EXPLORING THE POTENTIAL OF LARGE LANGUAGE MODELS FOR AUTOMATION IN TECHNICAL CUSTOMER SERVICE

Jochen Wulf, Jürg Meierhofer
Zurich University of Applied Sciences,
Technikumstrasse 81, 8401 Winterthur, Switzerland

**ABSTRACT**
**Purpose:** The purpose of this study is to investigate the potential of Large Language Models (LLMs) in transforming technical customer service (TCS) through the automation of cognitive tasks.
**Design/Methodology/Approach:** Using a prototyping approach, the research assesses the feasibility of automating cognitive tasks in TCS with LLMs, employing real-world technical incident data from a Swiss telecommunications operator.
**Findings:** Lower-level cognitive tasks such as translation, summarization, and content generation can be effectively automated with LLMs like GPT-4, while higher-level tasks such as reasoning require more advanced technological approaches such as Retrieval-Augmented Generation (RAG) or finetuning ; furthermore, the study underscores the significance of data ecosystems in enabling more complex cognitive tasks by fostering data sharing among various actors involved.
**Originality/Value:** This study contributes to the emerging theory on LLM potential and technical feasibility in service management, providing concrete insights for operators of TCS units and highlighting the need for further research to address limitations and validate the applicability of LLMs across different domains.

**KEYWORDS:** Technical Customer Service, Automation, LLM, AI

## 1. INTRODUCTION

Many firms struggle to provide a reliable technical customer service (TCS) with fast response times due to several challenges such as skilled labor shortage and information overload (Özcan et al., 2014). Large Language Models (LLMs) like OpenAI's GPT-4 are set to revolutionize TCS by providing efficient, and personalized support (Kanbach et al., 2023; Wulf & Meierhofer, 2023). They potentially handle high volumes of customer interactions, reducing the need for extensive human resources, and offer significant cost savings (Liu et al., 2023).

Non-generative AI has transformed contact centers, with AI-based tools automating customer service and sales, performing back-office tasks, and enabling remote monitoring, coaching, training, and scheduling (Doellgast et al., 2023). Practical evidence suggests that machine learning plays a crucial role by automating the resolution of TCS issues and requests, thereby minimizing business interruptions. It enhances the efficiency of TCS processes, such as incident and service request management, by automating manual tasks like ticket categorization and prioritization, thus reducing errors and improving overall process efficiency (Mönning et al., 2018).

While first academic studies gather evidence of the business impact of LLMs, the full potential of LLMs in transforming customer service is yet to be fully understood (Brynjolfsson et al., 2023). The purpose of this research is to identify cognitive tasks in TCS that may be automated. Furthermore, this research assesses feasibility in a real-world setting via a proof of concept.

## 2. THEORETICAL BACKGROUND
### 2.1 Automation of Cognitive Tasks with Large Language Models
An analysis of the academic literature on language modelling (Dasgupta et al., 2023; Liu et al., 2023; Radford et al., 2019; Wei et al., 2022; Zhao et al., 2023) yields five cognitive tasks that may be automated with LLMs in customer service contexts.

1) Translation and Correction: LLMs can translate text between languages or language modes by learning the underlying patterns and structures of different languages from the training data.
2) Summarization: LLMs can understand the relationships between different parts of the text and can generate a summary that accurately reflects the overall meaning of the original text.
3) Content Generation: LLMs can generate a wide range of content, from emails and social media posts to blog articles and stories.
4) Question Answering: In question answering the LLM either searches and uses the internal factual knowledge provided in the pre-training corpus or the external contextual data provided in the prompt to generate common sense answers to questions or instructions.
5) Reasoning: Complex reasoning, unlike common sense question answering, necessitates the comprehension and application of evidence and logic to reach conclusions. Typically, it involves a sequential reasoning process grounded in factual knowledge, culminating in the answer to a posed question.

### 2.2 Role of AI in Technical Customer Service
Prior research has shown that well-designed customer service practices enhance the cocreation of customer value (Winkler & Wulf, 2019). Setting up an efficient TCS is not a trivial task. It demands considerable expertise in designing and implementing operational processes, as well as in adopting the necessary technologies (Wulf & Winkler, 2020).

Several authors discuss the role of non-generative AI for TCS. Iparraguirre-Villanueva et al. (2023) review the literature on conversational bots and propose a technical architecture for bots in incident management. Chaturvedi and Verma (2023) discuss the transformative potential of AI in customer service, highlighting both its value-creating aspects like personalization and convenience, and its challenges such as privacy concerns and technology anxiety. Reinhard, Wischer et al. (2023) propose an AI-based conversational co-agent to assist in the onboarding of novice IT support agents, aiming to reduce job demand, augment problem-solving capabilities, and improve time-to-performance by considering cognitive load. Reinhard, Li et al. (2023) address the use of recommender systems in customer service. They present an analytics pipeline designed to improve data quality in service management. Using this pipeline, high-quality support tickets are extracted from a dataset of 60,000 real-life tickets, leading to better prediction performance in the instantiated recommender system compared to traditional methods.

The academic literature on the application of LLMs for customer service reflects a growing interest across various domains. Carvalho and Ivanov (2023) outline the profound impact likely on tourism through the integration of ChatGPT and other LLMs, such as improving front-of-house customer service and back-of-house operations efficiency. Potential risks are also addressed, suggesting that technological shifts often affect human resource roles, but may ultimately enhance the capabilities of tourism employees. Research on customer satisfaction in public services suggests the value of LLMs in analyzing online user feedback. Topic models, which can convert user opinions into actionable insights, have been proposed to help improve public service provision, with studies showing that the quality of staff interactions correlates strongly with user satisfaction (Kowalski et al., 2017). LLMs are recognized for their significance in question-answering and chatbot capabilities applied to healthcare, education, and customer service. The development of scalable clustering pipelines that fine-tune LLMs has been pivotal in surfacing user intentions from large volumes of conversational texts, enhancing the performance of data analysts and ultimately reducing the time needed to deploy chatbots (Chen & Beaver, 2022).

Reinhard et al. (2024) discuss how generative AI can lead to more efficient and higher-quality customer support. They identify five support activities that can be augmented with LLMs: assigning, referring & transferring, escalating, locating, adapting, generating, and retaining.

Brynjolfsson et al. (2023) empirically study the introduction of an LLM-based conversational assistant to 5,179 customer support agents and show a productivity increase of 14%, particularly benefiting novice and low-skilled workers. The AI tool also improved customer sentiment, reduced managerial intervention requests, and enhanced employee retention.

To summarize, existing studies indicate a significant potential for employing LLMs in TCS. However, these discussions are largely theoretical. The limited number of prototype and implementation studies only showcase a fraction of the potential benefits discussed. The analysis of the literature reveals a research gap in demonstrating the practicality and outlining the necessary conditions for automating cognitive tasks with LLMs in TCS.

## 3. RESEARCH METHODOLOGY

We use prototyping as research methodology, because prototyping in design research serves to transition abstract concepts into concrete forms, fostering clarity from vagueness (Barzilai & Ferraris, 2023; Camburn et al., 2017). Thus prototyping is well qualified to study the feasibility of theoretical approaches for the application of LLMs in TCS discussed in prior literature.

We implement proofs of concepts using data on technical incidents of a large telecommunications operator. The data consists of 1) a description of the customer incident, 2) a flow of messages discussing the problem and 3) a description of the solution. An example is shown in Table 1.

Table 1: Customer incident example

| Incident Description | *"I want to upgrade from [router old] to [router new]. Is it possible to upload the backup from the [router old] to the [router new], so that I don't have to manually transfer all configuration settings? Or is there another trick / migration procedure?"* |
|---|---|
| Problem Discussion (selection) | *"If you haven't made any special settings like port forwarding, then you don't need to do anything. When you connect the router (whether reset or not), you will be asked for your information on the login page. Once logged in, it will then load the previous data for WLAN and SIP (DECT or wired telephone) from the server onto [router new]. Thus, everything is as with the standard. Only then can you adjust the special parts or reload them via the locally stored file."* |
| Solution Description | *"I turned on and reset the [router new]. Then I connected the telephone line and waited until all data was received from [telco] and the LED was constantly white. Then I uploaded the previously performed backup of the [router old] from the local storage. Everything works perfectly as before (fixed IPs, Ports)."* |

For each cognitive task, we design dedicated prompts and issue these to the transformer-based GPT-4 chatbot provided by OpenAI (OpenAI et al., 2023). We use prompting good practices, such as Persona and Chain-of-Thought (White et al., 2023; Zhao et al., 2023). For validation, we manually compare the LLM output to the messages and solutions generated by the human support staff.

## 4. RESULTS

### 4.1 Text Correction and Translation

Language Models (LLMs) have the ability to convert text from one language or mode to another. They achieve this by understanding the patterns and structures of various languages, which they learn from the data they are trained on. The translation procedure requires the model to be given a prompt in the original language and then asked to generate a response in the desired language. The model's capacity to comprehend context and semantics enables it to produce translations of high quality. The following illustration (Figure 1) shows how GPT-4 transforms a basic email template provided in the prompt into a well-crafted customer email.

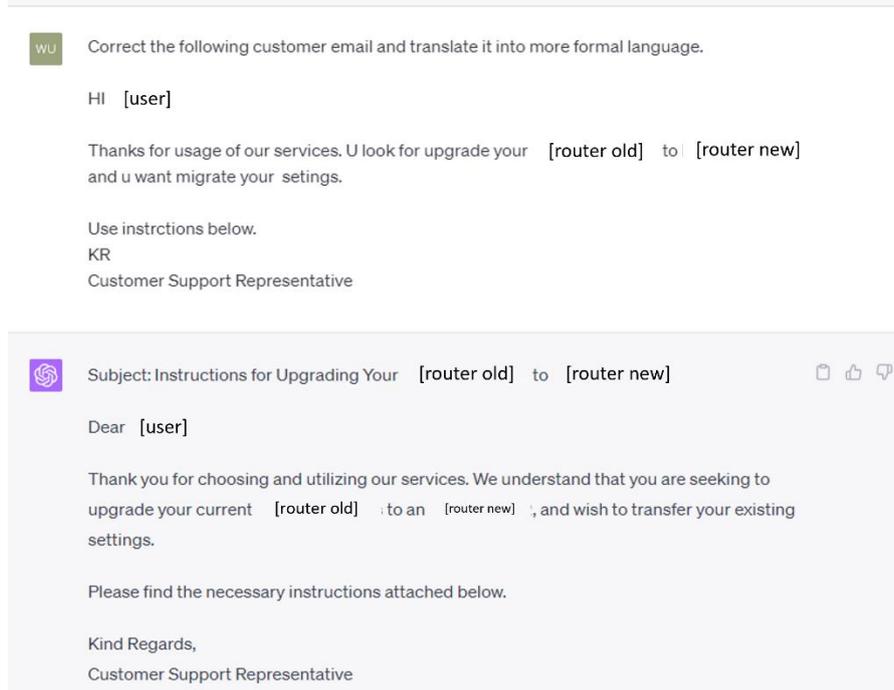

Figure 1: Translation example

## 4.2 Text Summarization

Contemporary LLMs are built upon the transformer architecture and utilize a feature known as the attention mechanism (Vaswani et al., 2017). This mechanism enables the model to consider the context of each word's usage. Consequently, the model can comprehend the connections between various sections of the text and produce a summary that precisely mirrors the original text's overall significance.

Text summarization will evolve into a crucial instrument for first-level staff to effectively glean vital data from previous customer interactions or incident reports. Following is a summarization example of an extensive customer dialogue concerning the router upgrade (Figure 2).

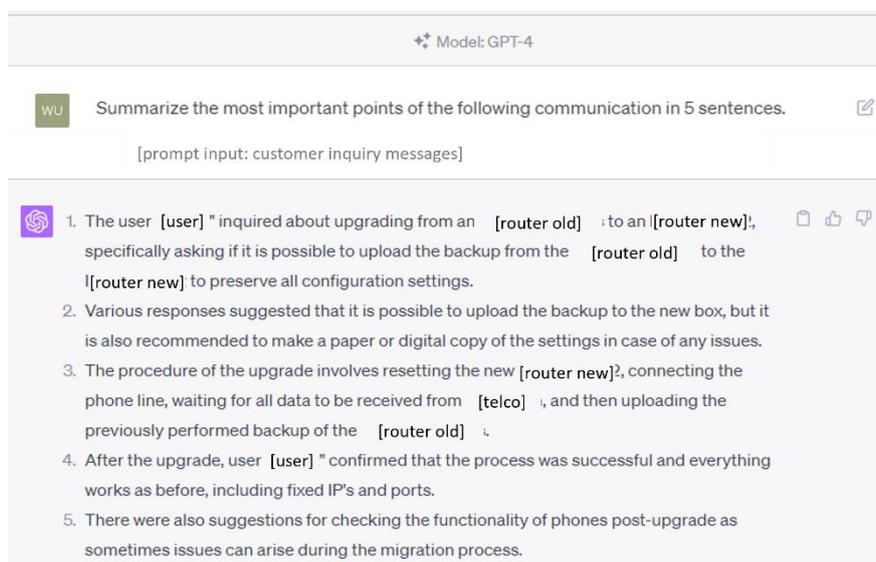

Figure 2: Summarization example

### 4.3 Content Generation

LLMs have the ability to produce diverse types of content, ranging from emails and posts on social media to blog entries and narratives. They can start with a cue or an initial idea, and then build the remaining content based on the patterns they've been trained on. This feature is beneficial in numerous fields, such as content marketing, creative writing, and beyond. The following example (Figure 3) illustrates how GPT-4 can craft an email for a customer, drawing from a previous sequence of messages related to a customer question.

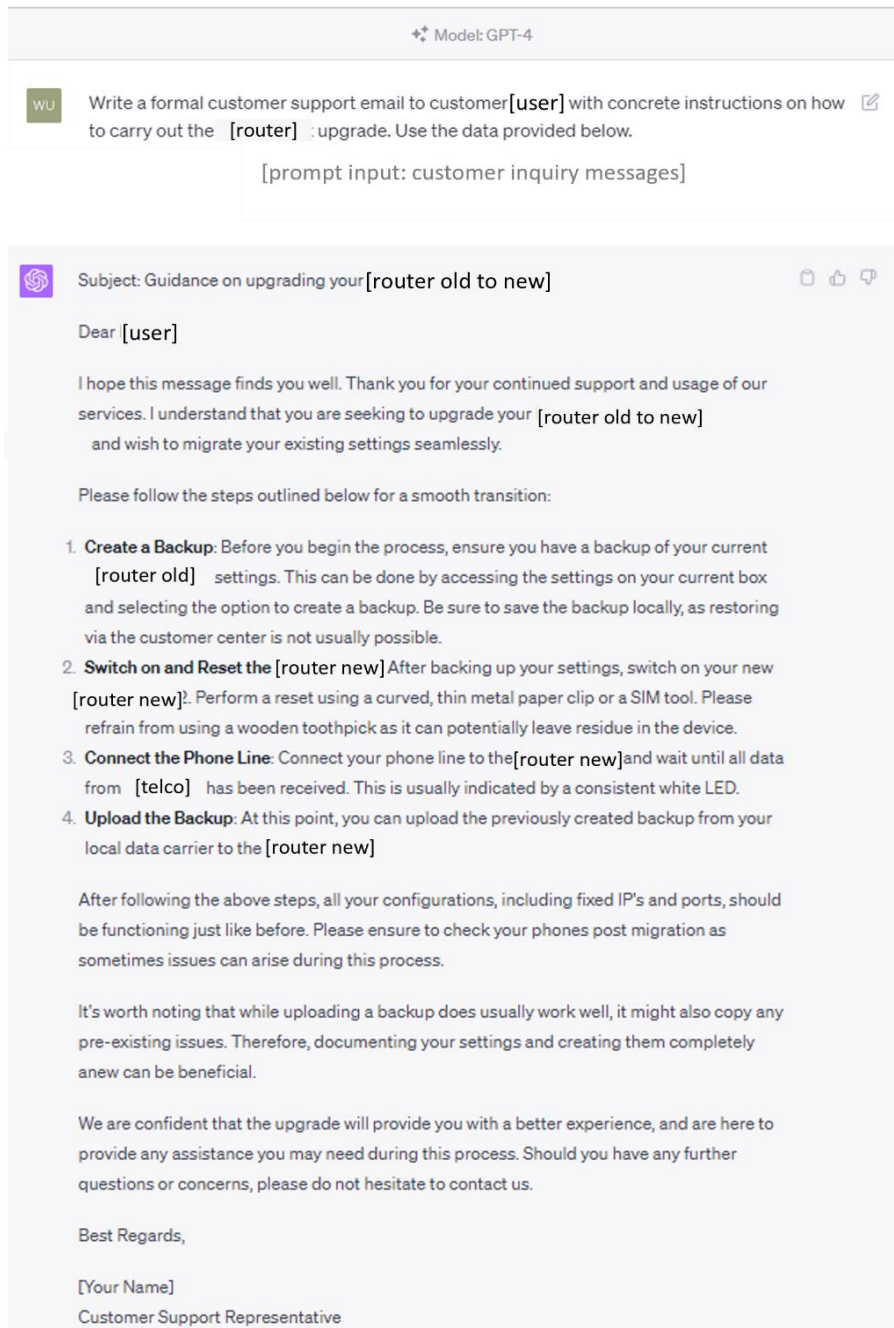

Figure 3: Example content generation

### 4.4 Question Answering

In question answering the LLM either searches and uses the internal factual knowledge provided in the pre-training corpus or the external contextual data provided in the prompt to generate commonsense answers to questions or instructions. In contrast to more complex reasoning tasks, question answering is limited to retrieving required and preexisting information from a large data set. The following example demonstrates the ability of GPT-4 to retrieve the solution for a customer problem from a larger problem-solution dataset (Figure 4). In this example, we handed over six different customer inquiry message flows, from which only one was relevant for the task at hand.

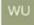

Figure 4: Example question answering

**4.5 Reasoning**
In the process of answering queries, the LLM either employs the inherent factual knowledge from the pre-training dataset or the external data from the given prompt to generate sensible responses to questions or directives. Unlike more intricate reasoning tasks, the scope of question answering is confined to extracting necessary and already available information from a vast dataset.
The subsequent example illustrates GPT-4's capability to extract a solution for a customer issue from a single problem-solution dataset (Figure 5).

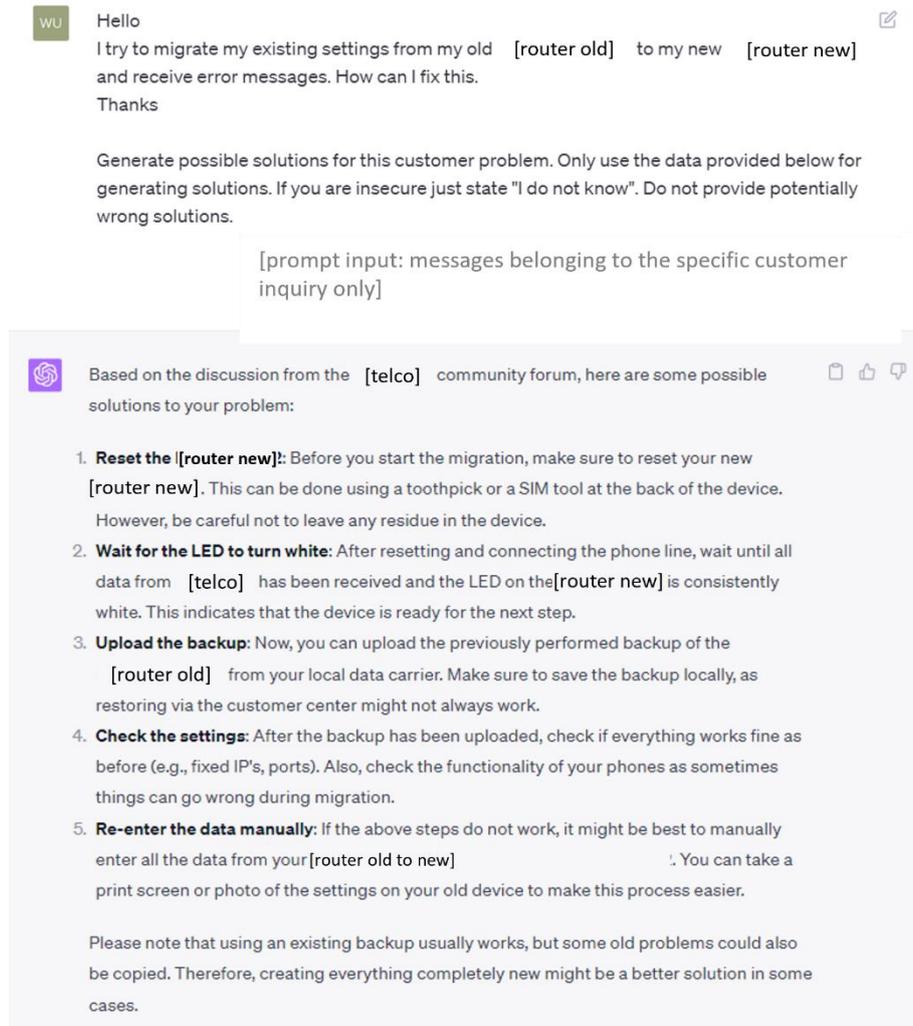

Figure 5: Example reasoning (with focused contextual data input)

In another instance (Figure 6), we employed the identical prompt as in Figure 5 but this time, we included six distinct customer query message flows. Out of these, only one was pertinent to the customer's issue. This situation is more akin to real-world cases where solutions need to be extracted from an extensive database of problems and their corresponding solutions. On this occasion, GPT-4 offered vague solution suggestions that wouldn't address the customer's issue. GPT-4 failed to pinpoint the exact problem and fetch the appropriate solution from the wider dataset. More complex reasoning tasks thus tend to produce hallucinations and require more complex and customized technological solution architectures.

> Hello
> I try to migrate my existing settings from my old [router old to new] and receive error messages. How can I fix this.
> Thanks
>
> Generate possible solutions for this customer problem. Only use the data provided below for generating solutions. If you are insecure just state "I do not know". Do not provide potentially wrong solutions.
>
> [prompt input: six different customer inquiry message flows]
>
> Based on the information gathered from the provided links, here are some possible solutions to your problem:
>
> 1. **Check Compatibility**: Ensure that the settings you are trying to migrate are compatible with the new [router new]. Some settings from the [router old] may not be supported by the newer model.
> 2. **Manual Configuration**: If the automatic migration of settings is not working, you may need to manually configure your settings on the new [router new]. This includes setting up your Wi-Fi network, firewall rules, and any other specific settings you had on your old device.
> 3. **Firmware Update**: Make sure that your new [router new] is running the latest firmware. Sometimes, issues with settings migration can be due to outdated firmware.
> 4. **Reset to Factory Settings**: If you're still having trouble, you might want to try resetting your [router new] to its factory settings and then try the migration process again.
> 5. **Contact [telco] Support**: If none of the above solutions work, it might be best to contact [telco]'s customer support. They can provide more specific guidance based on the exact error messages you're receiving.
>
> Remember, when setting up your new device, consider the optimal placement for your router. The placement can significantly affect the performance of your Wi-Fi network. The router should ideally be placed in a central location, where it can provide the best coverage for your devices. If you have a larger area to cover, you might want to consider using additional access points.
>
> Please note that these are general solutions based on the information provided in the links. The exact solution may vary depending on the specific error messages you're receiving.

Figure 6: Example reasoning (with broad contextual data input)

## 5. DISCUSSION

In summary our results show that lower-level cognitive tasks such as translation, summarization and content generation can be well automated with general purpose LLMs such as GPT-4. Higher level tasks such as reasoning, however, require more advanced technological approaches, such as Retrieval-Augmented Generation (RAG) or finetuning. RAG is a method that enhances the accuracy and credibility of large language models, particularly for knowledge-intensive tasks, by incorporating knowledge from external databases, allowing for continuous knowledge updates and integration of domain-specific information (Gao et al., 2023). LLM fine-tuning is the process of further training a pre-trained base LLM, or foundational model, for a specific task or knowledge domain (Lin et al., 2024). Depending on type and complexity of serviced equipment, task automation can be applied to and create value for onsite, front or back office TCS or even for customer self-service.

Apart from technological, there are also organizational challenges when it comes to realizing more complex cognitive tasks such as question answering and reasoning. These applications heavily rely on data sharing among the different actors involved in TCS, which must cooperate in a data ecosystem. Data Ecosystems can be described as intricate socio-technical systems where various participants engage and cooperate to discover, store, share, utilize, or repurpose data. These interactions not only stimulate innovation and generate value, but also pave the way for the emergence of new business models (S. Oliveira et al., 2019). Most often, TCS requires cooperation in an ecosystem of the following actors (Herterich et al., 2023):

- **Fellow customers:** The sharing of data about historic incidents and solutions among fellow customers of a technical service increases the overall incident-solution database on which an LLM operates. All customers would benefit from an improved question answering or reasoning in a self-service chatbot..
- **Technical product company:** It is the primary entity that produces technical goods or services. It is at the center of the data ecosystem and directly interacts with customers, providing technical support and services. It must share a rich set of data including product documentation, technical specifications, customer communication data, log data of technical events, and solution documentation. The manufacturing company benefits via a higher level of internal or customer-side automation of TCS with LLMs.
- **Service Desk Providers**: These entities handle customer inquiries and complaints. They share data about customer interactions, problem descriptions, and resolutions. This data can be used to train the LLM to handle similar issues in the future, reducing the workload on the service desk and improving customer satisfaction.
- **Field Service Providers**: These are the technicians who visit customer sites to resolve issues. They share data about the technical problems they encounter and their solutions. This data is invaluable for improving the LLM's ability to guide customers through troubleshooting steps, potentially avoiding the need for a site visit.
- **Suppliers of Technical Components**: These entities provide the parts used in the manufacturing process. They share data about component specifications, failure rates, and maintenance procedures. This data can help the LLM identify customer issues relating to component errors and facilitate the problem management for incidents at the second and third level.
- **IT Providers Enabling Industry 4.0 Solutions:** These entities provide the sensors, telecommunication networks, and software that form the basis for Industry 4.0 solutions. They share data about sensor readings, network performance, and software functionality. This data is crucial for the LLM to understand the real-time status of the product, predict potential issues, and suggest optimal operating conditions. By sharing this data, IT providers enable the LLM to provide more accurate and timely TCS.
- **Providers of Manufacturing Solutions**: These entities provide the machinery and processes used in manufacturing of a product. They share data about the production process and product components. This data can be used by the LLM to pinpoint customer issues relating to production errors or defective components. By sharing data, manufacturing solutions providers improve operational support for the products.

In conclusion, each actor in the ecosystem not only benefits individually from the data sharing but also contributes to the overall improvement of the LLM-based TCS.

### 6.CONCLUSION

With our results we contribute to the emerging theory on LLM potential and technical feasibility in service management. Further, we provide concrete insights for operators of TCS units. We demonstrate the automation of cognitive tasks with real-world examples.

However, it is important to acknowledge the limitations of this study to fully understand the scope and potential future directions of this research. The first limitation of this study is that it is based on technological prototypes developed using limited data. While these prototypes provide a proof of concept, they may not fully represent the complexities and variations of real-world scenarios. The study calls for further research involving large scale validation of technological feasibility. The prototypes need to be tested on a larger scale to ensure they can handle the volume and variety of data in a real-world setting. This would provide a more robust validation of LLM capabilities and its readiness for deployment.

Another limitation is the need for the usefulness of the technology in a technical customer service context to be empirically studied. While the technology may show promise in a controlled environment, its effectiveness in a practical setting, such as a TCS department, needs to be evaluated. This involves studying factors such as user acceptance, ease of integration with existing systems, and impact on service quality and efficiency.

The study is domain-specific, focusing on the telecommunications sector. Therefore, the results may not be generalized to other customer service domains without further research. Each domain has its unique characteristics and challenges, and a solution effective in one domain may not necessarily translate to success in another. Cross-domain studies are needed to explore the applicability and adaptability of LLMs for TCS across different sectors.

In conclusion, while the study has made significant strides in demonstrating LLM potential for TCS, these limitations highlight the need for further research. Addressing these limitations in future studies would provide a more comprehensive understanding of LLM potential and pave the way for its successful implementation in various domains.

**AUTHORS**

Jochen Wulf:
School of Engineering, Zurich University of Applied Sciences (ZHAW)
jochen.wulf@zhaw.ch

Jürg Meierhofer:
School of Engineering, Zurich University of Applied Sciences (ZHAW)
juerg.meierhofer@zhaw.ch